\documentclass{easychairmod}
\usepackage{amsfonts,amsmath,amssymb,mathtools} % for basic math stuff
\usepackage[only,llbracket,rrbracket,llparenthesis,rrparenthesis]{stmaryrd}
\usepackage{xcolor,listings}
\usepackage{tikz} %for diagrams

\usepackage{doc}

\definecolor{keywordcolor}{rgb}{0.7, 0.1, 0.1}   % red
\definecolor{tacticcolor}{rgb}{0.0, 0.1, 0.6}    % blue
\definecolor{commentcolor}{rgb}{0.4, 0.4, 0.4}   % grey
\definecolor{symbolcolor}{rgb}{0.0, 0.1, 0.6}    % blue
\definecolor{sortcolor}{rgb}{0.1, 0.5, 0.1}      % green
\definecolor{attributecolor}{rgb}{0.7, 0.1, 0.1} % red

\newtheorem{theorem}{Theorem}[subsection]

\newcommand{\simnot}{\mathord{\sim}}

\definecolor{darkgreen}{rgb}{0.18,0.54,0.34}
\definecolor{darkpink}{rgb}{0.75,0.25,0.5}

\makeatletter
\lst@InputCatcodes
\def\lst@DefEC{%
	\lst@CCECUse \lst@ProcessLetter
	^^80^^81^^82^^83^^84^^85^^86^^87^^88^^89^^8a^^8b^^8c^^8d^^8e^^8f%
	^^90^^91^^92^^93^^94^^95^^96^^97^^98^^99^^9a^^9b^^9c^^9d^^9e^^9f%
	^^a0^^a1^^a2^^a3^^a4^^a5^^a6^^a7^^a8^^a9^^aa^^ab^^ac^^ad^^ae^^af%
	^^b0^^b1^^b2^^b3^^b4^^b5^^b6^^b7^^b8^^b9^^ba^^bb^^bc^^bd^^be^^bf%
	^^c0^^c1^^c2^^c3^^c4^^c5^^c6^^c7^^c8^^c9^^ca^^cb^^cc^^cd^^ce^^cf%
	^^d0^^d1^^d2^^d3^^d4^^d5^^d6^^d7^^d8^^d9^^da^^db^^dc^^dd^^de^^df%
	^^e0^^e1^^e2^^e3^^e4^^e5^^e6^^e7^^e8^^e9^^ea^^eb^^ec^^ed^^ee^^ef%
	^^f0^^f1^^f2^^f3^^f4^^f5^^f6^^f7^^f8^^f9^^fa^^fb^^fc^^fd^^fe^^ff%
	^^^^0393^^^^2081^^^^2082^^^^2208^^^^03c3
	^^00}
\lst@RestoreCatcodes
\makeatother

\lstdefinestyle{lean}{
	inputencoding=utf8,
	extendedchars=true,
	belowcaptionskip=1\baselineskip,
	breaklines=true,
	mathescape=true,
	keywordstyle=[1]{\ttfamily\color{keywordcolor}},
	keywordstyle=[2]{\ttfamily\color{sortcolor}},
	keywordstyle=[3]{\ttfamily\color{tacticcolor}},
	keywordstyle=[4]{\ttfamily\color{attributecolor}},
	commentstyle=\color{blue}\textit,
	stringstyle=\color{pink}\ttfamily,
	basicstyle=\footnotesize\ttfamily,
	showstringspaces=false,
	identifierstyle={\color{black}}, 
	literate=*
	{|}{{{\color{darkgreen}$|$}}}{1}
	{?}{{{\color{darkgreen}$\to$}}}{1}
	{:}{{{\color{darkgreen}:}}}{1}
	{=}{{{\color{darkgreen}=\;}}}{1}
	{@}{{{\color{darkgreen}@}}}{1}
	{[}{{{\color{darkgreen}[}}}{1}
	{]}{{{\color{darkgreen}]}}}{1}
	{sigma}{{{\color{black}$\sigma$}}}{1}
	{Γ}{{$\Gamma$}}{1}
	{⊃}{{$\supset$}}{1}
	{.7}{{{\color{red}.7}}}{2},%
	morekeywords={
		import,open,reducible,noncomputable,notation,prefix,infix,local,attribute,
		example,definition,def,lemma,theorem,axiom,inductive,instance,structure,if,then,else,
		nat,bool,
		begin,end,by,match,with,cases,induction,fapply
	}
	morekeywords=[2]{Sort, Type, Prop},
	morekeywords=[3]{
		assumption,
		apply, intro, intros, allGoals,
		generalize, clear, revert, done, exact,
		refine, repeat, cases, rewrite, rw,
		simp, simp_all, contradiction,
		constructor, injection,
		induction,
	},
}

\newcommand{\nat}{\mathbb{N}}
\newcommand{\SV}{\mathrm{S5}}

\title{A Henkin-style completeness proof for the modal logic S5\thanks{The author wishes to thank Jeremy Avigad, Mario Carneiro, Rajeev Gor{\'e}, and Minchao Wu for helpful suggestions and invaluable conversations on the topic covered herein. An early version of this development was presented at the Lean Together 2019, Amsterdam, January 7--11, 2019. }}

\author{
Bruno Bentzen%\inst{1}
}

\institute{
  Carnegie Mellon University \\
  Pittsburgh, Pennsylvania, USA \\
  \email{b.bentzen@hotmail.com} }

\authorrunning{B. Bentzen}

\titlerunning{A Henkin-style completeness proof for the modal logic S5}

\begin{document}

\maketitle

\begin{abstract}
 \noindent 
  This paper presents a recent formalization of a Henkin-style completeness proof for the propositional modal logic S5 using the Lean theorem prover~\cite{de2015lean}. The proof formalized is close to that of Hughes and Cresswell~\cite{hughes1996new}, but the system, based on a different choice of axioms, is better described as a Mendelson system augmented with axiom schemes for K, T, S4, and B, and the necessitation rule as a rule of inference. The language has the false and implication as the only primitive logical connectives and necessity as the only primitive modal operator. The full source code is available online at \url{https://github.com/bbentzen/mpl/} and has been typechecked with Lean 3.4.2. 
\end{abstract}

\section{Introduction}
\label{sect:introduction}

A proof of the completeness theorem for a given logic conforms to the Henkin style when it applies nonconstructive methods to build models out of maximal consistent sets of formulas (possibly after a Henkin language extension) using the deductive system itself. Henkin-style completeness proofs for modal logics have been around for over five decades~\cite{negri2009kripke} but the formal verification of completeness with respect to Kripke semantics is comparatively recent. 

This paper presents a formalization of a Henkin-style completeness proof for the propositional modal logic S5 using the Lean theorem prover~\cite{de2015lean}. The proof is specific to S5, but, by forgetting the appropriate extra accessibility conditions (as described in~\cite{hughes1996new}), the technique we use can be applied to weaker normal modal systems such as K, T, S4, and B. The formalization covers the syntax and semantics of S5, syntactic and semantic deduction theorem, structural rules (weakening, contraction, exchange), the recursive enumerability of the language, and soundness and completeness results. In total, it has approximately 1,500 lines of code, but only two thirds of it is required for the completeness proof. The full source code is available online at \url{https://github.com/bbentzen/mpl/} and has been typechecked with Lean 3.4.2. 
At the time of writing, this is the latest stable release of Lean.

\subsection{Related work}
\label{sect:related}

The use of proof assistants in the mechanization of completeness proofs  in the context of Kripke semantics has been recently studied in the literature for a variety of formal systems.
Coquand~\cite{coquand2002formalised} uses ALF to give a constructive formal proof of soundness and completeness with respect to Kripke models 
for intuitionistic propositional logic with implication as the sole logical constant. 
Building on Coquand's work, a constructive completeness proof of Kripke semantics for intuitionistic logic with implication and universal quantification has been verified with Coq by Heberlin~and~Lee~\cite{herbelin2009forcing}. 
Also using Coq, Doczal~and~Smolka present a constructive formal proof of completeness with respect to Kripke semantics and decidability of the forcing relation for an extension of the basic modal logic K~\cite{doczkal2012constructive} and a variety of temporal logic~\cite{doczkal2014completeness}. 
In his formal verification of cut elimination for coalgebraic logics, Tews~\cite{tews2013formalizing} provides a formalization of soundness and completeness proofs covering many different logics, including modal logic K. 

However, to the best of our knowledge, the formalization of a Henkin-style completeness proof for propositional modal logic S5 proposed in this paper is the first of its kind.\footnote{
	In independent work done roughly at the same time the author first completed this formalization in 2018, Wu and Gor{\'e}~\cite{wu_et_al:LIPIcs:2019:11086} have described a formalization in Lean of modal tableaux for modal logics K, KT, and S4 with decision procedures with proofs of soundness and completeness. Also in 2018, but unknown to the author, From~\cite{Epistemic_Logic-AFP} formalized a Henkin-style completeness proof for system K in Isabelle. 
}
Our proof is close to that of Hughes and Cresswell~\cite{hughes1996new}, but given for a system based on a different choice of axioms. 
In Hughes and Cresswell's book, the basis of S5 is that of T plus an additional axiom. Here S5 is built on axiom schemes for K, T, S4 and B. This has the advantage that we can easily adapt the proof for different weaker systems. 
Another choice had to be made between using a deep or a shallow embedding for the formalization. Because our aim is metatheoretical, we use a deep embedding for the encoding of the syntax, as it allows us to prove metatheorems by structural induction on formulas or derivations. 

\subsection{Lean}
\label{sect:lean}

Lean is an interactive theorem prover developed principally by Leonardo de Moura and based on a version of dependent type theory known as the Calculus of Inductive Constructions~\cite{pfenning1989inductively,coquand1986calculus}. 
Theorem proving in Lean can be done by constructing proof terms directly as in Agda~\cite{agda}, by using tactics (imperative commands that describe how to construct a proof term) as in Coq~\cite{coq}, or by mixing them together in the same proof environment. 
Like most proof assistants, Lean also supports classical reasoning, 
which is employed in the formalization along with the declaration of noncomputable constructions. 
Finally, the formalization also presupposes a few basic results on data structures which are not in the standard library, so we make use of \texttt{mathlib}, the library of formalized mathematics for Lean~\cite{carneiro2018mathlib}. 

In the remainder of this paper, Lean code will be used to illustrate design choices and to give a broad overview of the proof method, but not to discuss the proof itself. Interested readers are encouraged to consult \texttt{completeness.lean}, the main file of the formalization where the crucial steps of the proof are given in detail. 
We shall also give an informal proof sketch of the completeness theorem using mathematical notation to convey the key ideas of the proof. 

%------------------------------------------------------------------------------
\section{Modal logic}
\label{sect:modal}

We start with the syntax and semantics of the modal logic we have formalized, namely, S5, the strongest of the five systems of modal logic proposed by Lewis and Langford~\cite{lewis1959symbolic}. In the next section, we give a proof sketch of completeness for the logic described here. 

\subsection{The language}
\label{sect:modallogic}

For simplicity, we shall work with a language which has implication ($\supset$) and the false ($\bot$) as the only primitive logical connectives, and necessity ($\Box$) as the only primitive modal operator. This choice of primitive operators gives rise to modal logics with a complex lattice structure~\cite{makinson1973warning}, but it can also be used to do intuitionistic modal logic if one wishes to do so. 
This language can be very conveniently defined using inductive types in Lean, in which case the parsing tree of a formula is always made explicit\textemdash in fact, knowing that a certain formula is well-formed amounts to knowing whether it is well-typed term of that type. 

\begin{lstlisting}[style=lean]
inductive form {$\sigma$ : nat} : Type
| atom : fin sigma $\to$ form
| bot  : form
| impl : form $\to$ form $\to$ form
| box  : form $\to$ form
\end{lstlisting}

\noindent Using one of the four constructors displayed above (\texttt{atom}, \texttt{bot}, \texttt{impl}, \texttt{box}) is the only way to construct a term of type \texttt{form}. 
The elimination rule of this type is precisely the principle of induction on the structure of a formula.

While newly-defined terms are always exhibited in Polish notation by default, 
Lean supports unicode characters and has an extensible parser which allows the declaration of customized prefix or infix notations for terms and types. 

\begin{lstlisting}[style=lean]
prefix   `#`      := form.atom
notation `$\bot$`       := form.bot _
infix    `$\supset$`       := form.impl
notation `$\simnot$`:40 p :=  form.impl p (form.bot _)
prefix   `$\Box$`:80    := form.box
prefix   `$\Diamond$`:80    := $\lambda$ p, $\simnot$($\Box$ ($\simnot$ p))
\end{lstlisting}

\noindent It is also possible to have notations for compound terms. Negation and possibility, for example, are encoded as definable logical and modal constants.
One of the benefits of having the false as primitive rather than negation is that the statement that a certain set of premises (context) is consistent need not be dependent on particular choices of variables. %For instance, compare
In other words, we can write \lstinline[style=lean]{$\Gamma \nvdash_{S5} \bot$} instead of \lstinline[style=lean]{$\Gamma \nvdash_{S5}$ p $\&~\simnot$p} to express that $\Gamma$ is consistent.

\subsection{Contexts}
\label{sect:ctx}

Contexts are just sets of formulas, so, since we already have a type of formulas, implementing contexts does not require extra work. However, as there is a variety of set-like objects in Lean (lists, multisets, finsets, sets etc.), there are many possible design choices. 

In an earlier version of the formalization, contexts were implemented using lists. A strong point of this approach is that one can reason by induction on contexts (lists are inductive types), allowing, among other things, a reduction of strong completeness to weak completeness via the deduction theorem~\cite{caiformalising}. Unfortunately, since lists must always be finite in size, there is no consistent way of reproducing maximal consistent extensions of contexts this way because every maximal consistent set is necessarily infinite. %\textemdash simply put, no Henkin-style completeness proof would be possible. %with that approach.

Contexts are now defined using sets, which are functions of type \texttt{A} $\to$ \color{darkpink}\texttt{Prop}\color{black}. 
Simply put, a set is just a predicate, a non-ordered collection that cannot contain duplicates (as a result, exchange and contraction are trivial structural rules of the proof system). 
For the purpose of stating more readable functions which look like logic textbook theorems, we have:

\begin{lstlisting}[style=lean]
def ctx : Type := set (form sigma)

notation `$\cdot$` := {}
notation $\Gamma$ `$_{_{\grave{}}}$` p := set.insert p $\Gamma$ 
\end{lstlisting}

\subsection{The proof system}
\label{sect:proofsystem}

We define a Hilbert-style axiomatic system, which is presented as a Mendelson system augmented with axiom schemes for K, T, S4, and B, and the necessitation rule as rule of inference. Such a proof system can be succinctly defined as follows in Lean:

\begin{lstlisting}[style=lean]
inductive prf : ctx $\sigma$ $\to$ form $\sigma$ $\to$ Prop
| ax {$\Gamma$} {p} (h : p$\in\Gamma$): prf $\Gamma$ p
| pl1 {$\Gamma$} {p q}        : prf $\Gamma$ (p $\supset$ (q $\supset$ p))
| pl2 {$\Gamma$} {p q r}      : prf $\Gamma$ ((p $\supset$ (q $\supset$ r)) $\supset$ ((p $\supset$ q) $\supset$ (p $\supset$ r)))
| pl3 {$\Gamma$} {p q}        : prf $\Gamma$ ((($\simnot$p) $\supset$ $\simnot$q) $\supset$ ((($\simnot$p) $\supset$ q) $\supset$ p)
| k {$\Gamma$} {p q}          : prf $\Gamma$ ($\Box$(p $\supset$ q) $\supset$ ($\Box$p $\supset$ $\Box$q))
| t  {$\Gamma$} {p}           : prf $\Gamma$ ($\Box$p $\supset$ p)
| s4 {$\Gamma$} {p}           : prf $\Gamma$ ($\Box$p $\supset$ $\Box\Box$p)
| b {$\Gamma$} {p}            : prf $\Gamma$ (p $\supset$ $\Box\Diamond$p)
| mp {$\Gamma$} {p q} (hpq: prf $\Gamma$ (p $\supset$ q)) (hp : prf $\Gamma$ p) : prf $\Gamma$ q
| nec {$\Gamma$} {p} (h : prf $\cdot$ p) : prf $\Gamma$ ($\Box$p)

notation $\Gamma$ ` $\vdash_{s5}$ ` p := prf $\Gamma$ p
notation $\Gamma$ ` $\nvdash_{s5}$ ` p := prf $\Gamma$ p $\to$ false
\end{lstlisting}

\noindent Note that the use of the empty context $\cdot$ in the definition of \texttt{nec} (the necessitation rule) is essential to the completeness proof described in this paper. We shall return to it soon when discussing the deduction theorem and semantic consequence relation. 

But before addressing those, it is worth pausing for a moment to give an illustration of how terms of the type of proofs defined above can be used to represent actual derivations in the modal logic S5. In particular, notice that the following term

\begin{lstlisting}[style=lean]
lemma box_dne {p : form $\sigma$} :
$\cdot \vdash_{\SV}$ ($\Box\simnot\simnot$p) $\supset$ ($\Box$p) :=
prf.mp (prf.k) (prf.nec (prf.dne))

\end{lstlisting}

\noindent translates as a proof tree of the following theorem 

\begin{center}
	
	\begin{tikzpicture}[
	edge from parent path={
		(\tikzparentnode\tikzparentanchor) 
		+(0pt,.5\tikzleveldistance)  
		-- (0pt,-.5\tikzleveldistance -| \tikzchildnode\tikzchildanchor) 
		-- +(2.20cm,0pt) 
		-- +(-2.20cm,0pt)
	},
	grow'=up,level distance=4ex,
	level/.style={sibling distance=15em/#1}]
	\node[label={[above right, xshift=4.75cm]:\scriptsize (mp)}] (A) {$\Box p \supset \Box \simnot\simnot p$}
	child { node [label={[above right, xshift=2.10cm]:\scriptsize (k)}] (B) {$\Box(p \supset \simnot \simnot p) \supset (\Box p \supset \Box \simnot\simnot p)$}
		child { node [label={[above right, xshift=1.40cm]:}] (C) {~}
	} }
	child { node [label={[above right, xshift=2.10cm]:{\scriptsize(nec)}}] (B) {$\Box(p \supset \simnot \simnot p)$}
		child { node [label={[above right, xshift=2.10cm]:{\scriptsize(dne)}}] (C) {$p \supset \simnot \simnot p$} 
			child { node [label={[above right, xshift=1.40cm]:}] (D) {~}
		} } 
	}
	;
	\end{tikzpicture}.
\end{center}

In some deductive systems for modal logic, the deduction theorem, which states that if a formula $q$ is deducible from a set of assumptions $\Gamma \cup \{p\}$ then the implication $p \supset q$ is deducible from $\Gamma$, does not hold. In this formulation, it can be proven by a straightforward induction on the structure of the proof tree or, what amounts to the same thing, by an application of the elimination rule of the type of proofs defined above.

\begin{lstlisting}[style=lean]
theorem deduction {$\Gamma$ : ctx $\sigma$} {p q : form $\sigma$} :
($\Gamma$ $_{_{\grave{}}}$ p $\vdash_{s5}$ q) $\rightarrow$ ($\Gamma$ $\vdash_{s5}$ p $\supset$ q) :=
begin
generalize eq : ($\Gamma ~_{_{\grave{}}}$ p) = $\Gamma$',
intro h,
induction h; subst eq,
{ repeat {cases h_h},
exact id,
{ exact mp pl1  (ax h_h) } },
{ exact mp pl1 pl1 },
{ exact mp pl1 pl2 },
{ exact mp pl1 pl3 },
{ apply mp,
{ exact (mp pl2 (h_ih_hpq rfl)) },
{ exact h_ih_hp rfl } },
{ exact mp pl1 k },
{ exact mp pl1 t },
{ exact mp pl1 s4 },
{ exact mp pl1 b },
{ exact mp pl1 (nec h_h) }
end
\end{lstlisting}

\noindent The proof of all cases follows from a simple application of modus ponens to axiom scheme (1), except for the modus ponens case that requires axiom scheme (2). 
The presence of the necessitation rule in the form of \texttt{nec} does not impose any difficulty in the proof because the context restriction in the antecedent gives us a stronger inductive hypothesis.

As noted in~\cite{hakli2012does}, whether the deduction theorem fails or not in modal logics which have the necessitation rule is a consequence of design choice. The necessitation rule, which informally says that if $p$ is a theorem of the system then so is $\Box p$, can be implemented in many ways. As can be seen above, we restrict its applicability to theorems by only allowing formulas that are derivable under no assumptions in the antecedent of \texttt{nec}. 
Another possibility is having a more general (i.e. unrestricted) formulation of the necessitation rule, which extends the range of application of \texttt{nec} to formulas that are derivable under assumptions. That would mean having a constructor such as \texttt{nec'} instead of \texttt{nec} in the formalization:

\begin{lstlisting}[style=lean,]
| nec' {$\Gamma$} {p} (h : prf $\Gamma$ p) : prf $\Gamma$ ($\Box$p)
\end{lstlisting}

\noindent while this rule can be useful and, indeed, the resulting system can be shown to be complete with respect to a global version of semantic consequence~\cite{popkorn1994first} (see section~\ref{sect:forcing}), the deduction theorem does not hold in the presence of it. 

\subsection{Semantics}
\label{sect:semantics}

\subsubsection{Kripke models}
\label{sect:model}

The semantics for S5 are given by Kripke semantics. A model $\mathcal{M}$ is a triple $\langle \mathcal{W,R}, v \rangle$ where
\begin{itemize}
	\item $\mathcal{W}$ is a non-empty set of objects called possible worlds;
	\item $\mathcal{R}$ is a binary, equivalence relation on possible worlds;
	\item $\mathsf{v}$ specifies the truth value of a formula at a world.  
\end{itemize}
Typically, there are no constraints on what kind of objects the members of $\mathcal{W}$ (possible worlds) should be. 
For our purposes, however, it is useful to let them be sets of formulas instead of arbitrary objects (there is no loss of generality in the resulting semantics). We thus have:

\begin{lstlisting}[style=lean]
def wrld ($\sigma$ : nat) := set (form $\sigma$)
\end{lstlisting}

\noindent Kripke models can be implemented as structures (inductive types with only one constructor). This can be done using the  \color{keywordcolor}\texttt{structure}\color{black}~command in Lean. 
In the following we define a 6-tuple composed of a domain, an accessibility relation, a valuation function, and reflexivity, symmetry and transitivity proofs for the given relation. 

\begin{lstlisting}[style=lean]
structure model :=
(wrlds  : set (wrld $\sigma$))
(access : wrld $\sigma$ $\to$ wrld $\sigma$ $\to$ bool)
(val    : fin $\sigma$ $\to$ wrld $\sigma$ $\to$ bool)
(refl   : $\forall$ w $\in$ wrlds, access w w = tt)
(symm   : $\forall$ w $\in$ wrlds, $\forall$ v $\in$ wrlds, access w v  = tt $\to$ access v w  = tt)
(trans  : $\forall$ w $\in$ wrlds, $\forall$ v $\in$ wrlds, $\forall$ u $\in$ wrlds, 
access w v  = tt $\to$ access v u  = tt $\to$ access w u  = tt)
\end{lstlisting}

The Boolean type \color{keywordcolor}\texttt{bool}\color{black}~is used in the formalization of truth values (i.e. either \texttt{tt}~or \texttt{ff}).

\subsubsection{Semantic consequence}
\label{sect:forcing}

We have a recursively defined forcing function which takes a model, a formula, and a world as inputs and returns a boolean value. %as its truth value. 
It can be defined by induction on the structure of the formula involved as follows:

\begin{lstlisting}[style=lean]
def forces_form (M : model) : form $\sigma$ $\to$ wrld $\sigma$ $\to$ bool
| (#p)    := $\lambda$ w, M.val p w
| (bot $\sigma$)  := $\lambda$ w, ff
| (p $\supset$ q) :=  $\lambda$ w, (bnot (forces_form p w)) || (forces_form q w) 
| ($\Box$p)     := $\lambda$ w, if ($\forall$ v $\in$ M.wrlds, w $\in$ M.wrlds $\to$ 
M.access w v = tt $\to$ forces_form p v = tt) then tt else ff
\end{lstlisting}

\noindent The definition is a routine adaptation of that found in traditional modal logic textbooks~\cite{hughes1996new}. Non-modal connectives are given truth-functionally and the necessity operator is defined by stating that a formula $\Box p$ is true at a world $w$ iff if $\mathcal{R}(w,v)$ then $p$ is true at $v$, for all $v \in \mathcal{W}$. 

This function can be extended to contexts in the obvious way: we say that a context is true at a world in a model if each formula of that context is true at that world and in that model. 

\begin{lstlisting}[style=lean]
def forces_ctx (M : model) ($\Gamma$ : ctx $\sigma$) : wrld $\sigma \to$ bool :=
$\lambda$ w, if ($\forall$ p, p $\in$ $\Gamma$ $\to$ forces_form M p w = tt) then tt else ff
\end{lstlisting}

A formula $p$ is a \textit{local} semantic consequence of a context $\Gamma$ iff, for all models $\mathcal{M}$ and for all worlds $w \in \mathcal{W}$, the fact that $\Gamma$ is true at $w$ in $\mathcal{M}$ implies that $p$ is true at $w$ in $\mathcal{M}$.

\begin{lstlisting}[style=lean]
inductive sem_csq ($\Gamma$ : ctx $\sigma$) (p : form $\sigma$) : Prop
| is_true (m : $\forall$ (M : model) (w : wrld $\sigma$), 
forces_ctx M $\Gamma$ w = tt $\to$ forces_form M p w = tt) : sem_csq

notation $\Gamma$ `$\vDash_{s5}$` p := sem_csq $\Gamma$ p
\end{lstlisting}

\noindent 
Say that a formula (context) is true in a model $\mathcal{M}$ if that formula (context) is true at all worlds in that model. 
A formula $p$ is a \textit{global} semantic consequence of a context $\Gamma$ iff, for any model $\mathcal{M}$, if $\Gamma$ is true in $\mathcal{M}$ then $p$ is true in $\mathcal{M}$. 
It is easy to see that local semantic consequence implies global semantic consequence, so the global version of completeness implies the local version. The converse does not hold. In particular, note that $\Box p$ is a global\textemdash but not a local\textemdash semantic consequence of $\{p\}$. The formalization includes local but not global semantic consequence. It is not hard to encode, but the global version of completeness is actually false for our proof system.\footnote{Unless \texttt{nec} is replaced with \texttt{nec'}. See~\cite{popkorn1994first} for an informal proof.}

%------------------------------------------------------------------------------
\section{The completeness theorem} 
\label{sect:completeness}

In this section we formalize a proof of completeness with respect to the proof system and semantics developed in the previous sections. First, we state the completeness theorem. Second, a general outline of the proof is presented. Next, we give an informal explanation of each individual proof step followed by a description of its respective implementation.

\begin{theorem}[Completeness]
	For every context $\Gamma$, any formula $p$ that follows semantically from $\Gamma$ is also derivable from $\Gamma$ in the modal logic S5. In symbols:
	$$ \Gamma \vDash_{\SV} p \Longrightarrow  \Gamma \vdash_{\SV} p$$
	That is, every semantic consequence is also a syntactic consequence in S5. 
\end{theorem}
The proof here requires full contraposition and it is thus non-constructive. More specifically, the idea is to assume that both $\Gamma \vDash_{\SV} p$ and $\Gamma \nvdash_{\SV} p$ hold, and then derive a contradiction using the syntax to build a model $\mathcal{M} = \langle \mathcal{W,R}, v \rangle$ (the canonical model) where $\Gamma$ is true but $p$ is false at a specific world in the domain. 
To complete the proof we shall need some additional lemmas, many of which can be easily proven by induction. 

We shall focus on sketching the formal argument for the following facts:

\begin{enumerate}
	\item $\Gamma \cup \{\simnot p\}$ has a maximal consistent extension $\Delta$ defined as follows: 
	\begin{align*}
	\Delta_0 :=& \Gamma \cup \{\simnot p\} \\
	\Delta_{n+1} :=&
	\begin{cases*}
	\Delta_n \cup \{\varphi_{n+1}\} & if $\Delta_n \cup \{\varphi_{n+1}\}$ is consistent  \\
	\Delta_n \cup \{\simnot \varphi_{n+1}\} & otherwise %\\
	\end{cases*} \\
	\Delta :=& \bigcup_{n \in \nat} \Delta_{n} %\{\Delta_{n}\}
	\end{align*} 
	(i.e. $\Delta$ is consistent, maximal and $\Gamma \cup \{\simnot p\} \subseteq \Delta$)
	
	\item There exists a canonical model where $p$ is true at $w$ iff $p \in w$; 
\end{enumerate}

\subsubsection{Maximal consistent sets}
\label{sect:extension}

We say that a context is maximal consistent if it is consistent and, moreover, for every formula expressible in the language, either it or its negation is contained in that context.

\begin{lstlisting}[style=lean]
def is_max ($\Gamma$ : ctx $\sigma$) := is_consist $\Gamma$ $\land$ ($\forall$ p, p $\in$ $\Gamma$ $\lor$ ($\simnot$p) $\in$ $\Gamma$)
\end{lstlisting}

\noindent As our language is countable, it is possible to construct each $\Delta_{n+1}$ using natural numbers to run through the set of all formulas, deciding whether or not a number's corresponding formula (when it exists) is consistent with $\Delta_n$ or not (another alternative is to generalize this construction to languages of arbitrary cardinalities using Zorn's lemma instead). 

The enumerability of the language is expressed using \texttt{encodable} types, which are constructively countable types. Essentially, a type $\alpha$ is encodable when it has an injection \lstinline[style=lean]{encode : $\alpha$ $\to$ nat} and a (partial) inverse \lstinline[style=lean]{decode : nat $\to$ option $\alpha$}.

\begin{lstlisting}[style=lean]
def insert_form ($\Gamma$ : ctx $\sigma$) (p : form $\sigma$) : ctx $\sigma$ :=
if is_consist ($\Gamma$ $_{_{\grave{}}}$ p) then $\Gamma$ $_{_{\grave{}}}$ p else $\Gamma$ $_{_{\grave{}}}$ $\simnot$p

def insert_code ($\Gamma$ : ctx $\sigma$) (n : nat) : ctx $\sigma$ :=
match encodable.decode (form $\sigma$) n with
| none   := $\Gamma$
| some p := insert_form $\Gamma$ p
end

def maxn ($\Gamma$ : ctx $\sigma$) : nat $\to$ ctx $\sigma$
| 0     := $\Gamma$
| (n+1) := insert_code (maxn n) n

def max ($\Gamma$ : ctx $\sigma$) : ctx $\sigma$ := 
$\bigcup$ n, maxn $\Gamma$ n	
\end{lstlisting}

\noindent Before proceeding any further, we must show that \texttt{max} $\Gamma$ is the maximal consistent extension of $\Gamma$, that is, that $\Gamma$ in contained in \texttt{max} $\Gamma$ and that \texttt{max} $\Gamma$ is maximal and consistent. 

First, we note that, for each \texttt{maxn} $\Gamma$ \texttt{n} of the family of sets, we have $\Gamma \subseteq$ \texttt{maxn} $\Gamma$ \texttt{n}. So $\Gamma$ must also be contained in their union, \texttt{max} $\Gamma$. This proof argument produces a term of type:

\begin{lstlisting}[style=lean]
lemma subset_max_self {$\Gamma$ : ctx $\sigma$} : 
$\Gamma \subseteq$ max $\Gamma$
\end{lstlisting}

\noindent Second, we observe that every formula must be in the enumeration somewhere, so suppose that the formula $p$ has index \texttt{i}. By the definition of \texttt{maxn} $\Gamma$ \texttt{i}, either $p$ or $\simnot p$ is a member of \texttt{maxn} $\Gamma$ \texttt{i}, so one of them is a member of \texttt{max} $\Gamma$. Thus, we have a term
\begin{lstlisting}[style=lean]
theorem mem_or_mem_max {$\Gamma$ : ctx $\sigma$} (p : form $\sigma$) :
p $\in$ max $\Gamma$ $\lor$ ($\simnot$p) $\in$ max $\Gamma$
\end{lstlisting}
\noindent Third, assume for the sake of contradiction that $\Gamma$ is consistent but \texttt{max} $\Gamma$ is not. By structural induction on the proof tree, we prove that there exists an \texttt{i} such that \texttt{maxn} $\Gamma$ \texttt{i} is inconsistent. However, each \texttt{maxn} $\Gamma$ \texttt{i} preserves consistency. This gives a function 
\begin{lstlisting}[style=lean]
lemma is_consist_max {$\Gamma$ : ctx $\sigma$} :
is_consist $\Gamma \rightarrow$ is_consist (max $\Gamma$)
\end{lstlisting}

The above proof sketches are implemented purely by unfolding definitions and inductive reasoning. They consist of approximately 150 lines of code in \texttt{completeness.lean}. 
There is even a one-line short case-reasoning proof that maximal consistent sets are closed under derivability: 

\begin{lstlisting}[style=lean]
lemma mem_max_of_prf {$\Gamma$ : ctx $\sigma$} {p : form $\sigma$} (h$_1$ : is_max $\Gamma$)
(h$_2$ : $\Gamma \vdash_{\SV}$ p) :  p $\in \Gamma$ :=
(h$_1$.2 p).resolve_right ($\lambda$ hn, h$_1$.1 (prf.mp (prf.ax hn) h$_2$))
\end{lstlisting}

\subsubsection{The canonical model construction}
\label{sect:canonicalmodel}

In this section we build the canonical model for $\SV$ from the deductive system itself. For that we specify a domain, an accessibility relation, a valuation function, and proofs of reflexivity, symmetry and transitivity for the accessibility relation.
We build the model as follows: 

\begin{itemize}
	\item $\mathcal{W}$ is the set of all maximal consistent sets of formulas;
	\item $\mathcal{R}(w,v)$ iff $\Box p \in w$ implies $p \in v$;
	\item $\textsf{v}(w,p)=1$ if $w\in \mathcal{W}$ and $p \in w$, for a propositional letter $p$.
\end{itemize}

\noindent For that to be a well-defined model, we must to show that $\mathcal{R}$ is an equivalence relation.

Reflexivity translates as follows: $\Box p \in w$ implies $p \in w$ for a given world $w \in \mathcal{W}$.
But this is easy because $w$ is closed under derivability (it is a maximal consistent set of formulas) and our proof system has modus ponens and axiom schema (t).

Proving symmetry requires more work. Given any worlds $w, v \in \mathcal{W}$, suppose first that $\Box \varphi \in w$ implies $\varphi \in v$ for all formulas $\varphi$, and suppose that $\Box p \in v$. We want to show that $p \in w$. Since $\Diamond\Box p \supset p$ is a theorem of S5
(see \texttt{syntax/lemmas.lean})
we just have to prove that $\Diamond \Box p \in w$, or, equivalently, that $\Box \simnot \Box p \notin w$. 
By contraposition on our initial hypothesis, it suffices to show that $\simnot \Box p \notin v$. But $\Box p \in v$ and $v$ is consistent.

For transitivity, we must show that 
$p \in u$, on the assumptions that $\Box p \in w$, that $\Box \varphi \in w$ implies $\varphi \in v$, and that $\Box \varphi \in v$ implies $\varphi \in u$, for any formula $\varphi$. 
In other words, we want to show that $\Box \Box p \in w$. But this follows from modus ponens and axiom scheme (s4). 

This model construction is represented by the Lean code

%/- the canonical model construction -/

\begin{lstlisting}[style=lean]
def domain ($\sigma$ : nat) : set (wrld $\sigma$) := {w | ctx.is_max w}

def unbox (w : wrld $\sigma$) : wrld $\sigma$ := {p | ($\Box$p) $\in$ w}

def access : wrld $\sigma$ $\to$ wrld $\sigma$ $\to$ bool :=
$\lambda$ w v, if (unbox w $\subseteq$ v) then tt else ff

def val : fin $\sigma$ $\to$ wrld $\sigma$ $\to$ bool :=
$\lambda$ p w, if w $\in$ domain $\sigma$ $\land$ (#p) $\in$ w then tt else ff
\end{lstlisting}

\begin{lstlisting}[style=lean]
lemma mem_unbox_iff_mem_box {p : form $\sigma$} {w : wrld $\sigma$} : 
p $\in$ unbox w $\leftrightarrow$ ($\Box$p) $\in$ w :=
$\langle$ id, id $\rangle$
\end{lstlisting}

\noindent What is here called \texttt{unbox} is a set operation which takes a set of formulas $w$ as an input and returns the set of formulas $p$ such that $\Box p$ is a member of $w$. 

A rather trivial but still quite useful lemma about this operation is that if $p$ is deducible from \texttt{unbox} $w$ then actually $\Box p \in w$. It can be proved by structural induction on the derivation using the necessitation rule. The proof argument is straightforward but, because we shall need this fact in the next section, it is convenient to have the formal proof presented here: 

\begin{lstlisting}[style=lean]
lemma mem_box_of_unbox_prf {p : form $\sigma$} {w : wrld $\sigma$}
(H : w $\in$ domain $\sigma$) : (unbox w $\vdash_{s5}$ p) $\to$ ($\Box$p) $\in$ w :=
begin
generalize eq : unbox w = $\Gamma$',
intro h, induction h; subst eq,
{ assumption },
repeat { apply ctx.mem_max_of_prf H,
apply prf.nec,
apply prf.pl1 <|> apply prf.pl2 <|> apply prf.pl3 },
{ apply ctx.mem_max_of_prf H,
refine prf.mp (prf.ax _) (prf.ax (h_ih_hp rfl)),
exact (ctx.mem_max_of_prf H) 
(prf.mp prf.k (prf.ax (h_ih_hpq rfl))) },
{ apply ctx.mem_max_of_prf H,
exact prf.nec prf.k },
{ apply ctx.mem_max_of_prf H,
exact prf.nec prf.t },
{ apply ctx.mem_max_of_prf H,
exact prf.nec prf.s4 },
{ apply ctx.mem_max_of_prf H,
exact prf.nec prf.b },
{ apply ctx.mem_max_of_prf H,
apply prf.nec (prf.nec h_h) }
end
\end{lstlisting}

\subsubsection{Truth and membership}
\label{sect:fullproof}

In order to prove completeness, we first show that truth is membership in the canonical model, that is, that a formula is true at a world in the canonical model iff it is a member of that world:

\begin{lstlisting}[style=lean]
lemma form_tt_iff_mem_wrld {p : form $\sigma$} :
$\forall$ (w $\in$ domain $\sigma$), (forces_form model w p) = tt $\leftrightarrow$ p $\in$ w
\end{lstlisting}

\noindent where \texttt{model} is the canonical model defined in the previous section. To prove this, we use induction on the structure of the formula $p$. So, we will show that the result holds when $p$ is a propositional letter $\# p$, the false $\bot$, an implication $p \supset q$ or a necessary proposition $\Box p$.

In the proof mechanization, we use the \texttt{\color{tacticcolor}induction} tactic in the tactic mode. 
This tactic produces four goals (as displayed below). %followed by a proof sketch.

\begin{lstlisting}[style=lean]
case form.atom
$\sigma$ : $\mathbb{N}$,
p : fin $\sigma$
$\vdash$ $\forall$ (w : wrld $\sigma$),
w $\in$ domain $\sigma \rightarrow$ forces_form model w #p = tt $\leftrightarrow$ #p $\in$ w)
\end{lstlisting}

\noindent The first goal is a base case. 
If $p$ is a propositional letter the biconditional statement is clearly true, because, by definition of the valuation function, this is what it means for a propositional letter to be true at a world in the canonical model. 

\begin{lstlisting}[style=lean]
case form.bot
$\sigma$ : $\mathbb{N}$,
$\vdash$ $\forall$ (w : wrld $\sigma$),
w $\in$ domain $\sigma \rightarrow$ forces_form model w $\bot$ = tt $\leftrightarrow$ $\bot$ $\in$ w)
\end{lstlisting}

\noindent Proving the second goal is easy. Both conditional statements are vacuously true in both directions for the false, because $w$ is consistent and, by definition, the false cannot be true at a world in a model. 

\begin{lstlisting}[style=lean]
case form.impl
$\sigma$ : $\mathbb{N}$,
p q : form $\sigma$,
hp : 
$\forall$ (w : wrld $\sigma$),
w $\in$ domain $\sigma \rightarrow$ forces_form model w p = tt $\leftrightarrow$ p $\in$ w),
hq : 
$\forall$ (w : wrld $\sigma$),
w $\in$ domain $\sigma \rightarrow$ forces_form model w q = tt $\leftrightarrow$ q $\in$ w)
$\vdash$ $\forall$ (w : wrld $\sigma$),
w $\in$ domain $\sigma \rightarrow$ forces_form model w (p $\supset$ q) = tt $\leftrightarrow$ (p $\supset$ q) $\in$ w)
\end{lstlisting}

\noindent For the third goal, we assume the inductive hypothesis that the result holds for $p$ and $q$, and then show that either $p$ is false or $q$ is true at $w$ in the canonical model iff $p \supset q \in w$. The proof follows from case reasoning and the closure under derivability of possible worlds.

\begin{lstlisting}[style=lean]
case form.box
$\sigma$ : $\mathbb{N}$,
p : form $\sigma$,
hp : 
$\forall$ (w : wrld $\sigma$),
w $\in$ domain $\sigma \rightarrow$ forces_form model w p = tt $\leftrightarrow$ p $\in$ w)
$\vdash$ $\forall$ (w : wrld $\sigma$),
w $\in$ domain $\sigma \rightarrow$ forces_form model w ($\Box$p) = tt $\leftrightarrow$ ($\Box$p) $\in$ w)
\end{lstlisting}

\noindent To prove the fourth goal, we begin by assuming the inductive hypothesis for $p$. If $w$ is a world, and, if it is a maximal consistent set of formulas, then, by unfolding the definition of truth of a formula at a world in a model, the biconditional statement becomes

\begin{lstlisting}[style=lean]
$\vdash$ ($\forall$ (v : wrld $\sigma$),
v $\in$ model.wrlds $\rightarrow$ w $\in$ model.wrlds $\rightarrow$ model.access w v = tt $\rightarrow$ forces_form model w p = tt) $\leftrightarrow$ ($\Box$p) $\in$ w)
\end{lstlisting}

In the forwards direction, we assume that $\Box p$ is true at $w$ in the canonical model and that $\simnot\Box p \in w$. But then, by lemma \texttt{mem\_box\_of\_unbox\_prf}, the context \texttt{unbox} $w \cup \{ \simnot p \}$ is consistent and can be extended to a maximal consistent set (i.e. a world in the domain). It is accessible to $w$ because \texttt{unbox} $w \subseteq$ \texttt{max} (\texttt{unbox} $w \cup \{ \sim p \}$), so $p$ should be true at $w$. But $p \notin$ \texttt{max} (\texttt{unbox} $w \cup \{\sim p\})$ because it is consistent.

For the backwards direction, assume that $\Box p \in w$. Given a maximal consistent set of formulas $v$ and assuming that $\Box \varphi \in w$ then $\varphi \in v$ for all $\varphi$, we have to show that $p$ is true at $v$ in the model.  By the inductive hypothesis, however, it suffices to show that $p \in v$, but this follows from $\Box p \in w$.

\subsubsection{The completeness proof}
\label{sect:fullproof}

We now complete our proof by putting together all the above pieces into 24 lines of code.
Since we know by hypothesis that $\Gamma\nvdash_{\SV} p$, it follows that $\Gamma \cup \{\simnot p\}$ is consistent\textemdash otherwise, if the false were deducible from it, we would have a contradiction by double negation elimination.

\begin{lstlisting}[style=lean]
lemma consist_not_of_not_prf {$\Gamma$ : form $\sigma$} {p : form $\sigma$} :
($\Gamma$ $\nvdash_{s5}$ p) $\rightarrow$ is_consist ($\Gamma$ $_{_{\grave{}}}$ $\simnot$p) :=
$\lambda$ hnp hc, hnp (mp dne (deduction hc))
\end{lstlisting}

\noindent Now assuming that $\Gamma\vDash_{\SV} p$, the basic idea for deriving the contradiction is that, as \texttt{max} $\Gamma \cup \{\simnot p\}$ is a world in the canonical model, and each formula $\varphi \in \Gamma$ is true at that world, $\Gamma$ is true as well. 
Clearly, $p$ is not consistent with  $\Gamma \cup \{\simnot p\}$, so $p \notin$ \texttt{max} $\Gamma \cup \{\simnot p\}$, meaning that $p$ must be false at that world. 

This allows us to prove the following theorem

\begin{lstlisting}[style=lean]
theorem completenss {$\Gamma$ : form $\sigma$} {p : form $\sigma$} :
($\Gamma$ $\vDash_{s5}$ p) $\rightarrow$ $\Gamma$ $\vdash_{s5}$
\end{lstlisting}

%------------------------------------------------------------------------------
\section{Conclusion}
\label{sect:conclusion}

We presented a computer formalization of the completeness of the modal logic S5 using a Henkin-style argument, as given in traditional textbooks such as~\cite{hughes1996new}.
An interesting direction for further work would be to formally verify the global version of completeness for the alternative presentation of S5 obtained by replacing \texttt{nec} with \texttt{nec'}. 
Another natural step for continuing the work presented here is to present completeness proofs for different modal logics. 

\vspace{5mm}
\noindent \textbf{Acknowledgments} \;  %The author is also indebted to ... for helpful comments on an earlier draft of this paper.
Work supported in part by the AFOSR grant FA9550-18-1-0120. Any opinions, findings and conclusions or recommendations expressed in this material are those of the author(s) and do not necessarily reflect the views of the AFOSR. 

\label{sect:bib}
\bibliographystyle{plain}
\bibliography{ref}

\begin{thebibliography}{10}

\bibitem{caiformalising}
Leran Cai, Ambrus Kaposi, and Thorsten Altenkirch.
\newblock {Formalising the Completeness Theorem of Classical Propositional
  Logic in Agda (Proof Pearl)}.
\newblock URL: \url{https://akaposi.github.io/proplogic.pdf}, 2015.

\bibitem{carneiro2018mathlib}
Mario Carneiro.
\newblock {The Lean 3 Mathematical Library (mathlib)}.
\newblock URL: \url{https://robertylewis.com/files/icms/Carneiro_mathlib.pdf},
  2018.
\newblock International Congress on Mathematical Software.

\bibitem{coquand2002formalised}
Catarina Coquand.
\newblock A formalised proof of the soundness and completeness of a simply
  typed lambda-calculus with explicit substitutions.
\newblock {\em Higher-Order and Symbolic Computation}, 15(1):57--90, 2002.
\newblock URL: \url{https://doi.org/10.1023/A:1019964114625}.

\bibitem{coquand1986calculus}
Thierry Coquand and G{\'e}rard Huet.
\newblock The calculus of constructions.
\newblock {\em Information and Compututation}, 76(2-3):95--120, 1988.
\newblock URL: \url{https://hal.inria.fr/inria-00076024/document}.

\bibitem{de2015lean}
Leonardo de~Moura, Soonho Kong, Jeremy Avigad, Floris Van~Doorn, and Jakob von
  Raumer.
\newblock The lean theorem prover (system description).
\newblock In A.~Felty and A.~Middeldorp, editors, {\em International Conference
  on Automated Deduction}, pages 378--388, Cham, 2015. Springer.
\newblock URL: \url{https://doi.org/10.1007/978-3-319-21401-6_26}.

\bibitem{doczkal2012constructive}
Christian Doczkal and Gert Smolka.
\newblock Constructive completeness for modal logic with transitive closure.
\newblock In C.~Hawblitzel and D.~Miller, editors, {\em International
  Conference on Certified Programs and Proofs}, pages 224--239, Berlin,
  Heidelberg, 2012. Springer.
\newblock URL: \url{https://doi.org/10.1007/978-3-642-35308-6_18}.

\bibitem{doczkal2014completeness}
Christian Doczkal and Gert Smolka.
\newblock {Completeness and decidability results for CTL in Coq}.
\newblock In G.~Klein and R.~Gamboa, editors, {\em International Conference on
  Interactive Theorem Proving}, pages 226--241, Cham, 2014. Springer.
\newblock URL: \url{https://doi.org/10.1007/978-3-319-08970-6_15}.

\bibitem{Epistemic_Logic-AFP}
Asta~Halkjær From.
\newblock Epistemic logic.
\newblock {\em Archive of Formal Proofs}, October 2018.
\newblock \url{https://devel.isa-afp.org/entries/Epistemic_Logic.html}, Formal
  proof development.

\bibitem{hakli2012does}
Raul Hakli and Sara Negri.
\newblock Does the deduction theorem fail for modal logic?
\newblock {\em Synthese}, 187(3):849--867, 2012.
\newblock URL: \url{https://doi.org/10.1007/s11229-011-9905-9}.

\bibitem{herbelin2009forcing}
Hugo Herbelin and Gyesik Lee.
\newblock {Forcing-based cut-elimination for Gentzen-style intuitionistic
  sequent calculus}.
\newblock In Ono H., Kanazawa M., and de~Queiroz~R., editors, {\em
  International Workshop on Logic, Language, Information, and Computation},
  pages 209--217, Berlin, Heidelberg, 2009. Springer.
\newblock URL: \url{https://doi.org/10.1007/978-3-642-02261-6_17}.

\bibitem{hughes1996new}
George~Edward Hughes and Max~J Cresswell.
\newblock {\em A new introduction to modal logic}.
\newblock Psychology Press, 1996.

\bibitem{lewis1959symbolic}
Clarence~Irving Lewis and Cooper~Harold Langford.
\newblock {Symbolic Logic. The Century Company, New York, 1932}, 1959.

\bibitem{makinson1973warning}
David Makinson.
\newblock A warning about the choice of primitive operators in modal logic.
\newblock {\em Journal of philosophical logic}, 2(2):193--196, 1973.
\newblock URL: \url{https://www.jstor.org/stable/30226058}.

\bibitem{negri2009kripke}
Sara Negri.
\newblock Kripke completeness revisited.
\newblock {\em Acts of Knowledge: History, Philosophy and Logic: Essays
  Dedicated to G{\"o}ran Sundholm}, pages 247--282, 2009.
\newblock URL: \url{https://www.mv.helsinki.fi/home/negri/gkcrev.pdf}.

\bibitem{agda}
Ulf Norell.
\newblock {Dependently typed programming in Agda}.
\newblock In P.~Koopman, R.~Plasmeijer, and D.~Swierstra, editors, {\em
  International School on Advanced Functional Programming}, pages 230--266,
  Berlin, Heidelberg, 2008. Springer.
\newblock URL: \url{https://doi.org/10.1007/978-3-642-04652-0_5}.

\bibitem{pfenning1989inductively}
Frank Pfenning and Christine Paulin-Mohring.
\newblock {Inductively defined types in the Calculus of Constructions}.
\newblock In {\em International Conference on Mathematical Foundations of
  Programming Semantics}, pages 209--228. Springer, 1989.

\bibitem{popkorn1994first}
Sally Popkorn.
\newblock {\em First steps in modal logic}.
\newblock Cambridge University Press, 1994.

\bibitem{coq}
The~Coq project.
\newblock The coq proof assistant.
\newblock URL: \url{http://www.coq.inria.fr}, 2017.

\bibitem{tews2013formalizing}
Hendrik Tews.
\newblock {Formalizing cut elimination of coalgebraic logics in Coq}.
\newblock In D.~Galmiche and D.~Larchey-Wendling, editors, {\em International
  Conference on Automated Reasoning with Analytic Tableaux and Related
  Methods}, pages 257--272, Berlin, 2013. Springer.
\newblock URL: \url{https://doi.org/10.1007/978-3-642-40537-2_22}.

\bibitem{wu_et_al:LIPIcs:2019:11086}
Minchao Wu and Rajeev Gor{\'e}.
\newblock {Verified Decision Procedures for Modal Logics}.
\newblock In J.~Harrison, J.~O'Leary, and A.~Tolmach, editors, {\em 10th
  International Conference on Interactive Theorem Proving (ITP 2019)}, pages
  31:1--31:19, Dagstuhl, Germany, 2019. Schloss Dagstuhl--Leibniz-Zentrum fuer
  Informatik.
\newblock URL: \url{https://doi.org/10.4230/LIPIcs.ITP.2019.31}.

\end{thebibliography}

\end{document}